\documentclass[pra,twocolumn,showpacs,aps,superscriptaddress,floatfix]{revtex4}
\usepackage{graphicx}

\begin{document}

\title{ Quantum-classical transition for an analog of double-slit
experiment in complex collisions: Dynamical decoherence in quantum
many-body systems}

\author{L.~Benet}%
\affiliation{Instituto de Ciencias F\'{\i}sicas, Universidad Nacional
  Aut\'onoma de M\'exico (UNAM), 62210--Cuernavaca, Mor., Mexico}%
\author{L.T.~Chadderton}%
\affiliation{Atomic and Molecular Physics Laboratary, RSPhysSE,
  Australian National University, Canberra ACT 0200, Australia}%
\author{S.Yu.~Kun}%
\affiliation{Facultad de Ciencias, Universidad Aut\'onoma del Estado
  de Morelos (UAEM), 62209--Cuernavaca, Mor., Mexico}%
\affiliation{Nonlinear Physics Center and Department of Theoretical
  Physics, RSPhysSE, Australian National University, Canberra ACT
  0200, Australia}%
\author{Wang~Qi}%
\affiliation{Institute of Modern Physics, Chinese Academy of Sciences,
  Lanzhou 730000, China}%

\date{\today}

\begin{abstract}
  We study coherent superpositions of clockwise and anti-clockwise
  rotating intermediate complexes with overlapping resonances formed
  in bimolecular chemical reactions. Disintegration of such complexes
  represents an analog of famous double-slit experiment. The time for
  disappearance of the interference fringes is estimated from
  heuristic arguments related to fingerprints of chaotic dynamics of a
  classical counterpart of the coherently rotating complex. Validity
  of this estimate is confirmed numerically for the H+D$_2$ chemical
  reaction. Thus we demonstrate the quantum--classical transition in
  temporal behavior of highly excited quantum many-body systems in the
  absence of external noise and coupling to an environment.
\end{abstract}


\maketitle

The famous double-slit experiment~\cite{1} provides the most vivid
demonstration of quantum coherent superpositions. These manifest
themselves in interference fringes of the intensity due to the
interference between the matter waves emerging from different slits.
The double-slit experiments have successfully demonstrated the wave nature
of, {\sl e.g.} electrons, neutrons, atoms, small molecules, noble gas
clusters and fullerenes~\cite{1}. A process of converting coherent
superpositions into classical sum of intensities manifests itself in
the disappearance of the interference fringes. This fundamental physical
process of the emergence of classical dynamics from the quantum-mechanical
description is referred to as the quantum--classical transition (QCT).

There are two possible roots to describe QCT. The first one is
decoherence~\cite{2} due to the external noise or coupling to the
environment. This mechanism of QCT results from a non-unitary
evolution of the system. The QCT due to the decoherence for the
double-slit experiment with fullerenes has been
demonstrated~\cite{3}. On the contrary, dynamical decoherence~\cite{4}
describes the QCT without coupling to environment and only due to the
intrinsic unitary evolution of a pure quantum state~\cite{5}. This
process, unlike decoherence~\cite{2}, describes the QCT on a finite
time scale shorter than Heisenberg time, which diverges in the
macroscopic limit~\cite{4}.

In this communication we address the problem of quantum--classical
transition in the absence of any coupling to the environment,
revealing an effect analogous to dynamical decoherence~\cite{4,5}. We
focus on the QCT in a temporary quantum evolution. This problem is
motivated by the work~\cite{5}, where time-integration appeared to be
a precondition for the quantum-classical transition. However, without
disappearance of the interference fringes at fixed moment of time the
QCT is clearly incomplete. This rises the question whether dynamical
decoherence~\cite{4} can lead to a QCT in a way that decoherence
does~\cite{2}.

Instead of a single-particle problem~\cite{5}, we consider coherent
superpositions of clockwise and anti-clockwise rotating many-body
intermediate complexes (IC) with strongly overlapping resonances. Such
IC can be created in atomic cluster collisions, bimolecular chemical
reactions and heavy-ion collisions. The physical picture of rotational
wave packets and their interference, originally revealed for heavy-ion
collisions~\cite{6,7,8}, has been strongly supported by numerical
calculations for, {\sl e.g.} the H+D$_2$~\cite{9}, F+HD~\cite{10} and
He+H$_2^+$~\cite{11} state-to-state chemical reactions. The many-body
aspect is of primary importance for the QCT since the macroscopic
world consists of complex systems. We demonstrate here that isolated
quantum systems can undergo a QCT and we obtain the characteristic
time for such transition.

We consider the spin-less collision partners in the entrance ($a$) and
exit ($b$) channels. The energy fluctuating $S$-matrix elements are
taken in the pole form~\cite{12}. This is a good approximation in the
regime of overlapping resonances of the IC, $\Gamma /D\gg 1$, where
$\Gamma$ is total width of resonance levels and $D$ is average level
spacing of the IC, provided $n_{ch}\gg\Gamma/D$ with $n_{ch}$ being a
number of open channels~\cite{12}. The high excitation energy of IC,
the smallness of the average level spacing of the IC, the strong
overlap of the resonance levels and the condition $n_{ch}\gg 1$ imply
that we are deeply in the semi-classical region. In this regime we
expect the quantum evolution to reveal fingerprints of classical
dynamics of the classical counterpart of the system. We shall see that
this is indeed the case.

Under the above stated conditions the time power spectrum for
the time-delayed collision has the form
$\Pi(t,\theta)=H(t)\exp(-t/t_{lt})|{\cal P}(t,\theta )|^2 $, where
\begin{eqnarray}
{\cal P}(t,\theta )&=& \frac{1}{N^{1/2}}\sum_{J}(2J+1)W(J)^{1/2}
\exp(i\Phi J)P_J(\theta ) \nonumber\\
&&\quad \sum_\mu{\tilde c}_\mu^J\exp(-iE_\mu^J t/\hbar ).
\label{eq1}
\end{eqnarray}
Here, $t$ is the time, $\theta $ is the scattering angle, $J$ is the
total spin of the IC, and $H(t)$ is the Heaviside step function. The
deflection angle $\Phi$ is given by the first $J$ derivative of a sum
of the potential phase shifts in the entrance and exit channels, and
the $P_J(\theta)$ are Legendre polynomials. In the $\Pi (t,\theta)$,
the factor $\exp(- t/t_{lt})$ accounts for the finite life-time,
$t_{lt}=\hbar /\Gamma$, of the IC. Therefore its energy levels have a
finite width, $\hbar/t_{lt}$, corresponding to the continuous
spectrum. On the other hand, even though the resonance levels are
strongly overlapping, $P(t,\theta )=|{\cal P}(t,\theta )|^2$ describes
the time evolution in discrete spectrum. Note that while $P(t,\theta
)$ is the probability, except for the omitted factor $\exp(-
t/t_{lt})$, for the IC to decay into an angle $\theta$ at the moment
$t$, it also describes the time evolution of the angular orientation
of the IC~\cite{13}. In Eq.~(\ref{eq1}), ${\tilde c}_\mu^J=
c_\mu^J/[\overline{(c_\mu^J)^2}{~}]^{1/2}$,
$c_\mu^J=\gamma_\mu^{Ja}\gamma_\mu^{Jb}-
\overline{\gamma_\mu^{Ja}\gamma_\mu^{Jb}}{~~}$ ($\overline{c_\mu^J} =
0$), $\gamma_\mu^{Ja(b)}$ and $\gamma_\nu^{J^\prime a(b)}$ are real
partial width amplitudes, and $E_\mu^J$, $E_\nu^{J^\prime}$ are the
resonance energies. The overlines stand for averaging over resonance
states. The number of resonance states included in the sum in
Eq.~(\ref{eq1}) is $N={\cal I}/D\gg 1$, where the energy interval
${\cal I}$ should be sufficiently long to resolve the interference
fringes~\cite{14}. In Eq.~(\ref{eq1}), $W(J)= \langle |\delta
S^J(E)|^2 \rangle$ is the average partial reaction probability taken
in the Gaussian form, $\exp[-(J-I)^2/d^2]$, where $I$ is the average
total spin of the IC and $d$ is the $J$-window width.

The intensity $|{\cal P}(t,\theta )|^2$ includes the sum
$\frac{1}{N}\sum_{\mu\nu}{\tilde c}_\mu^J {\tilde c}_\nu^{J^\prime }
\exp[i(E_\nu^{J^\prime}-E_\mu^J ) t/\hbar ]$. We calculate these sums by
performing first a partial $(\mu,\nu )$-summation with $(E_\mu^J
-E_\nu^{J^{\prime} })$ fixed within an uncertainty of about a few
units of $D$. Then, we get instead of the product ${\tilde
  c}_\mu^J{\tilde c}_\nu^{J^\prime}$ its average over resonances,
$\overline{{\tilde c}_\mu^J{\tilde c}_\nu^{J^\prime}}$. The latter is
given~\cite{6} by
\begin{equation}
  \overline{{\tilde c}_\mu^J{\tilde c}_\nu^{J^\prime}} = 
  \frac{(1/\pi)D\beta |J-J^\prime|}{ [E_\mu^J-E_\nu^{J^\prime} - 
    \hbar\omega ( J-J^\prime )]^2+\beta^2(J-J^\prime )^2}\ ,
  \label{eq_ref6}
\end{equation}
where $\beta \gg D$ is the phase relaxation width and $\omega$ is the
angular velocity of the coherent rotation of the highly-excited
IC. The spin diagonal contribution is evaluated by performing first a
partial $(\mu,\mu^\prime )$-summation with fixed $(\mu -\mu^\prime )$
value. This yields $\overline{{\tilde c}_\mu^J{\tilde
    c}_{\mu^\prime}^{J}}= \delta_{\mu\mu^\prime}$. The above
approximation is justified if the time for formation of the many-body
eigenstates of the effective Hamiltonian of the IC originated from the
intramolecular energy relaxation due to, {\sl e.g.} a strong
anharmonic coupling between the normal vibrational modes, is faster
than a characteristic rotation time of the IC. The diagonal
approximation, for the states with the same $J$-values, is the
standard assumption in the theory of quantum chaotic
scattering~\cite{15}. The employed approximations imply that instead
of the products ${\tilde c}_\mu^J{\tilde c}_\nu^{J^\prime}$ we can use
their average over resonances. This substitution does not violate the
time quasiperiodicity of $|{\cal P}(t,\theta )|^2$, which is
characteristic of a unitary evolution, since it is still given by a
discrete superposition of a finite number of Fourier components with
discrete frequencies $(E_\mu^J-E_\nu^{J^\prime})/\hbar$.

We first calculate the sums 
\begin{equation}
  \frac{1}{N}\sum_{\mu\nu} \overline{{\tilde c}_\mu^J {\tilde c}_\nu^{J^\prime }}
  \exp[i(E_\nu^{J^\prime}-E_\mu^J ) t/\hbar ] 
\label{num3}
\end{equation}
in the macroscopic limit $\hbar\to 0$. Since $D\propto \hbar^f$,
where $f0$ is a number of degrees of freedom, the macroscopic limit
corresponds to a continuous spectrum approximation. Changing from the
summation to integration in Eq.~(\ref{num3}) we obtain
\begin{equation}
 \exp[-i\omega
t (J-J^\prime)] \exp[-\beta t |J-J^\prime|/\hbar]. 
\label{eq4}
\end{equation}
Therefore the continuous spectrum approximation results in an
irreversible decay of the spin off-diagonal correlations implying a
non-unitary non-periodic time evolution. In order to preserve a
unitary time-periodic evolution we must evaluate the sums~(\ref{num3})
for a discrete spectrum. We do so using first the equidistant spectrum
approximation. Such an approximation is justified for $t\ll
2\pi\hbar/D$, when the spectrum is not resolved and thus the detailed
spectral properties do not affect the time behavior. Then, for ${\cal
  I}\gg\hbar\omega,\beta$, the original sums~(\ref{num3}) are reduced
to
\begin{displaymath}
  (1/\pi )D\beta |J-J^\prime
  |\sum_{k=-\infty}^\infty\frac{\exp(-ikDt/\hbar ) }{ [kD-\hbar\omega
  (J-J^\prime )]^2+\beta^2(J-J^\prime )^2 }
\end{displaymath}
with $k$ being integers. We calculate these sums using the Poisson
summation formula and obtain
\begin{eqnarray}
    \sum_{M=-\infty }^{\infty } \exp[-2\pi
  (\beta /D)|J-J^\prime ||M-Dt/2\pi\hbar |\,] \nonumber\\
   \exp[i2\pi (\hbar\omega/D)(J-J^\prime )(M-Dt/2\pi\hbar )],
\label{eq_sum}
\end{eqnarray}
where $M$ are integers. The above expression is periodic in time, with
the period $2\pi\hbar/D$, resulting in a periodicity of $|{\cal
  P}(t,\theta )|^2$ with this same period. However, we are interested
in the time evolution of the IC with strongly overlapping resonances
corresponding to relatively short time intervals $0\le t\ll 2\pi\hbar
/D$. Then, the above sums are dominated by the $M=0$ term since, for
$\beta\gg D$, the $M\neq 0$ terms are exponentially small and can be
neglected. Then, the $M=0$ term equals the expression ~(\ref{eq4}) and
we observe that, for $t< \pi\hbar/D$, the continuous and equidistant
spectrum approximation yield identical results. Yet, while the
continuous spectrum approximation violates a unitary periodic in time
evolution, the equidistant spectrum approximation obviously leads to a
unitary time-periodic evolution. We have also calculated the
sums~(\ref{num3}) numerically with $E_\mu^J$ and $E_\nu^{J^\prime}$
discrete spectra being uncorrelated for $J \ne J^\prime$ and each
having the universal spectral fluctuations of random matrix theory
(Wigner--Dyson statistics~\cite{guhr98}). We have found that, for
$D\ll \beta$ and $t\ll 2\pi \hbar/D$, these sums are accurately given
by the expression~(\ref{eq4}), {\sl i.e.}, by the $M=0$ term in
Eq.~(\ref{eq_sum}). Again, this is because for $t\ll 2\pi\hbar /D$ the
spectrum is not resolved and the sums~(\ref{num3}) are insensitive to
the detailed spectral properties. Altogether, for $t\ll 2\pi \hbar/D$,
we have
\begin{eqnarray} 
P(t,\theta) \propto  \sum_{JJ^\prime}(2J+1)(2J^\prime +1)
[W(J)W(J^\prime)]^{1/2} \nonumber\\
 P_J(\theta)P_{J^\prime}(\theta) \exp[i(\Phi-\omega t)(J-J^\prime )  
-\beta |J-J^\prime |t/\hbar ] .
\label{eq2}
\end{eqnarray}
Obviously, the above result can also be obtained within the continuous
spectrum approximation which is valid only for $t\le \hbar/D$ when the
spectrum is not resolved.  Clearly the appearance of the decaying
factor $\exp (-\beta |J-J^\prime |t/\hbar )$ for $t < \hbar/D$ does
not violate the unitary character of evolution of the IC. The problem
of experimental determination of $P(t,\theta )$ has been discussed
in~\cite{14,16}.

We employ the near-side far-side decomposition~\cite{11} of the
collision amplitude, $ {\cal P}(t,\theta)={\cal P}^{(-)}(t,\theta
)+{\cal P}^{(+)}(t,\theta )$, so that $ P(t,\theta )\propto |{\cal
  P}^{(+)}(t,\theta )|^2 +|{\cal P}^{(-)}(t,\theta )|^2 
+2{\rm Re} [{\cal P}^{(+)}(t,\theta ){\cal P}^{(-)}(t,\theta )^\ast ]
$. The amplitudes ${\cal P}^{(\pm)}(t,\theta )$ have the form of
Eq.~(\ref{eq1}) but with $ Q_J^{(\pm )}(\theta)=(1/2)[P_J(\theta )\mp
(2i/\pi)Q_J(\theta )] $ instead of $P_J(\theta )$.  Here $Q_J(\theta
)$ are Legendre functions of the second kind of degree $J$ and the
traveling Legendre functions have the semi-classical asymptotics $
Q_J^{(\pm )}(\theta)\sim [1/2\pi J\sin\theta ]^{1/2}\exp\{\pm
i[(J+1/2)\theta -\pi /4]\} $.

It is expected that the quantum superpositions originate from the
interference between the ${\cal P}^{(-)}(t,\theta )$ and ${\cal
  P}^{(+)}(t,\theta )$ in a close vicinity of $\theta =0,\pi$, where
the wave packets overlap. However, for these angle ranges, the
asymptotic form of $Q_J^{(\pm )}(\theta)$ is not valid. Yet,
$t_{q-cl}$ may be conjectured from the time dependence of the
near-side and far-side intensities, $|{\cal P}^{(\pm)}(t,\theta )|^2$,
for the intermediate angle interval, where the asymptotic form of
$Q_J^{(\pm )}(\theta)$ is an accurate approximation.

We calculate $|{\cal P}^{(\pm)}(t,\theta )|^2$ using the same
approximations as those employed for the calculation of $P(t,\theta
)$. The $|{\cal P}^{(\pm)}(t,\theta )|^2$ have the form of
Eq.~(\ref{eq2}) but with the asymtotics of $Q_J^{(\pm )}(\theta)$
instead of $P_J(\theta )$. Calculations analogous to those of~\cite{6}
yield that, for $t\ll\hbar /d\beta$, $|{\cal P}^{(\pm)}(t,\theta )|^2$
describe Gaussian rotational wave packets having a constant width
(angular dispersion) $\simeq 1/d$. Thus, for $t\ll\hbar /d\beta$,
there is no spreading of the wave packets and the memory about their
width at $t=0$ remains. This is characteristic of purely quantum
integrable-like evolution while the classical counterpart of the
many-body IC is clearly chaotic. On the contrary, for $t\gg \hbar
/d\beta$, the width of the wave packets becomes much greater than that
at $t=0$ and the memory about the angular dispersion at $t=0$ is lost.
In addition, for $t\gg \hbar /d\beta$ and $\beta t/\hbar >1$, $|{\cal
  P}^{(\pm)}(t,\theta )|^2$ approach a uniform distribution
exponentially fast~\cite{6}, $|{\cal P}^{(\pm)}(t,\theta )|^2\propto \exp[2\cos
(\Phi\pm \theta -\omega t)/\exp(\beta t/\hbar )]$. We conjecture that
this long-time behavior may be considered as fingerprint of chaotic
dynamics of the classical counterpart of the IC. Since the crossover
from the integrable-like to the chaotic-like dynamics of the quantum
evolution occurs between $t_1\ll\hbar /\beta d$ and $t_2\gg\hbar
/\beta d$, the characteristic time for the QCT can be evaluated as
$t_{q-cl}\sim \hbar/\beta d$. But then another feature of the QCT
should be the disappearance of the interference between waves
originating from the decay of rotating in opposite directions IC. This
transformation of the quantum superposition into an incoherent sum of
the intensities should also happen at $t_{q-cl}\sim \hbar/\beta d$.

\begin{figure}[t]
\includegraphics[angle=270,width=8.0cm]{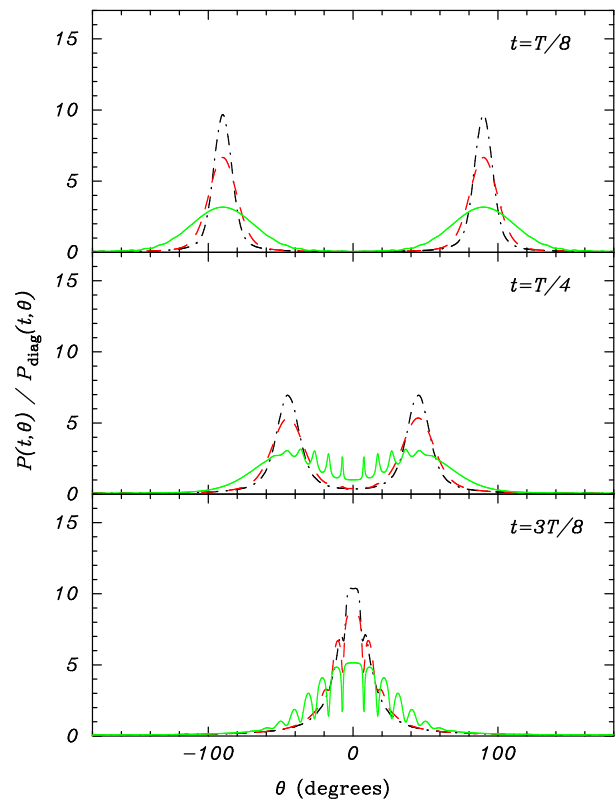}
\caption{\label{fig1} (Color online) Time power spectra related to the
  H+D$_2(v_i=0,j_i=0 )\to$HD$(v_f=3,j_f=0 )$+D chemical reaction for
  different moments of time. Solid lines correspond to $d=2$ and
  $\beta=0.003$ eV, dashed lines $d=5$ and $\beta=0.003$ eV, and
  dashed dotted lines $d=10$ and $\beta=0.003$ eV.}
\end{figure}

\begin{figure}[t]
\includegraphics[angle=270,width=8.0cm]{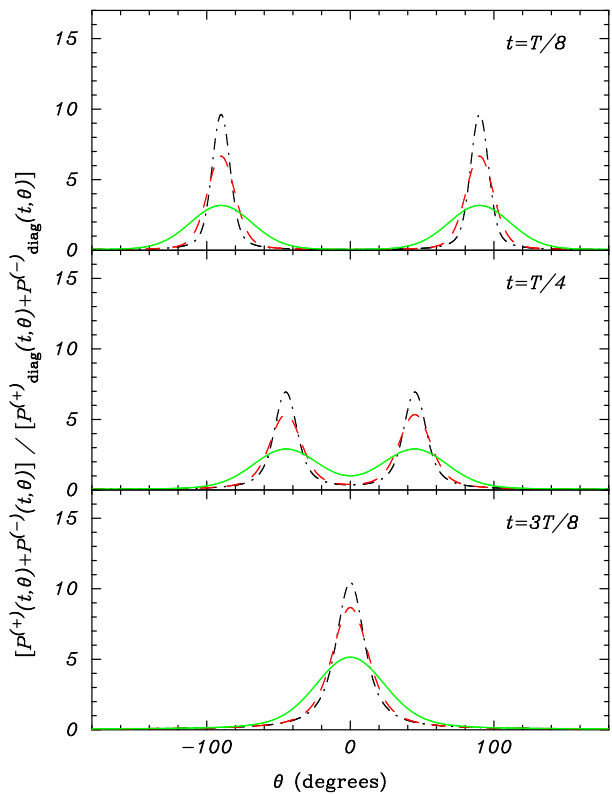}
\caption{\label{fig2} (Color online) Classical limits for the time
  power spectra presented in Fig.~1 (see text).  }
\end{figure}

To test this we use Eq.~(\ref{eq2}) to describe rotational wave
packets and their interference revealed~\cite{9} for the chemical
reaction H+D$_2(v_i=0,j_i=0 )\to$HD$(v_f=3,j_f=0 )$+D, where $v_i,j_i$
and $v_f,j_f$ are vibrational and rotational quantum numbers for the
initial and final states, respectively. In this reaction, the wave
packets likely originate from interference of the overlapping
resonances of the IC~\cite{17}. From the numerical
calculations~\cite{9} we deduce $\Phi=135^\circ$ and
$\hbar\omega=0.045$ eV. It was found that the main contribution to the
time-delayed reaction mechanism comes from the total angular momenta
$J=15-20$. Accordingly, we choose $I=18$. To reproduce the width of
the wave packets, $\simeq 25^\circ$, at an early stage, when
$\theta\simeq \Phi$, we take $d=2$. To account for the strong
interference contrast in forward direction we choose $\beta=0.003$~eV.

In Fig.~\ref{fig1} we present $P(t,\theta )/P_{diag}(\theta)$, where
$P_{diag}(\theta )$ is a time-independent quantity given by the
diagonal $J=J^\prime$ terms in Eq.~(\ref{eq2}). The reason for this
scaling is that amplitudes of individual wave packets are strongly and
abruptly enhanced in the vicinity of $\theta=0,\pi$ due to the
azimuthal symmetry of the problem ~\cite{8}. However,
$P_{diag}(\theta)$ has the similar enhancement for $\theta\sim 0,\pi$,
so that the quantity $P(t,\theta )/ P_{diag}(\theta)$ permits us to
probe interference between the wave packets whose amplitudes depend
smoothly on $\theta$. Another advantage of this scaling is that
$P_{diag}(\theta)$ corresponds to the very quick phase-relaxation
$(\hbar/\beta\to 0 )$ yielding the universal limit of the theory of
quantum chaotic scattering~\cite{15}. In this limit, wave packets are
uniformly spread and their interference is completely destroyed due to
the absence of the spin off-diagonal contributions. Thus, a deviation
of $P(t,\theta )/ P_{diag}(\theta)$ from a constant unity is a
quantitative measure of the deviation of the collision process from
the universal limit of the quantum chaotic scattering theory.

In Fig.~\ref{fig1} we present $P(t,\theta )/ P_{diag}(\theta)$ for
three moments of time. One can see that the wave packets, initially
oriented along $\theta=\Phi,2\pi -\Phi$, rotate towards each other.
Interference fringes are produced when the wave packets start to
overlap. For $d=2$, $\beta=0.003$ eV we have $t_{q-cl}\simeq
\hbar/\beta d=220$ fs. This time is longer by about a factor of 6.5
than the longest time $3T/8$ in Fig.~\ref{fig1}, where
$T=2\pi/\omega=90$ fs. We observe a strong contrast of the
interference fringes. Now we decrease $t_{q-cl}$ by taking $d=5$ and
keeping $\beta$ the same. Then, for $t=3T/8\simeq 0.38\times
t_{q-cl}$, the interference fringes are suppressed but still visible
in the bottom panel of Fig.~\ref{fig1}. We further decrease $t_{q-cl}$
by taking $d=10$ with the value of $\beta$ unchanged. Then, for
$t=3T/8\simeq 0.77\times t_{q-cl}$, the interference fringes
practically disappear manifesting a transformation of the coherent
superpositions into the incoherent mixture of the intensities. Thus,
we have confirmed the validity of our estimate for the $t_{q-cl}$.
This estimate has also been confirmed by calculations where we have
increased $\beta$ keeping $d=2$ fixed.

To test that the interference fringes do originate from the
interference between the ${\cal P}^{(+)}(t,\theta )$ and ${\cal
  P}^{(-)}(t,\theta )$ we calculate $P^{(+)}(t,\theta )+
P^{(-)}(t,\theta )$, where $P^{(\pm )}(t,\theta )=|{\cal P}^{(\pm
  )}(t,\theta )|^2$. We employ the same approximations used in
calculating $P(t,\theta )$, Eq.~(\ref{eq2}). We obtain $P^{(\pm
  )}(t,\theta )$ having the same form as Eq.~(\ref{eq2}) but with
$Q_J^{(\pm )}(\theta)Q_{J^\prime}^{(\pm )}(\theta)^\ast $ instead of
$P_J(\theta )P_{J^\prime }(\theta )$. In Fig.~\ref{fig2} we present
$[P^{(+)}(t,\theta )+ P^{(-)}(t,\theta )]/[P_{diag}^{(+)}(t,\theta )+
P_{diag}^{(-)}(t,\theta )]$ for the same values of parameters as used
in Fig.~\ref{fig1}. The time independent quantity
$[P_{diag}^{(+)}(t,\theta )+ P_{diag}^{(-)}(t,\theta )]$ contains the
diagonal $J=J^\prime$ terms only. The reason for this scaling is the
same as that employed in Fig.~\ref{fig1}. In addition, $Q_J^{(\pm
  )}(\theta)$ have a logarithmic divergence at $\theta=0,\pi$
resulting in unphysical increase of $P^{(\pm )}(t,\theta)$ in a close
vicinity of $\theta=0,\pi$. The presentation in Fig.~\ref{fig2} allows
to scale out these undesirable features enabling a meaningful
comparison with Fig.~\ref{fig1}. We see that the interference fringes
in Fig.~\ref{fig1} do originate from the interference between the
${\cal P}^{(+)}(t,\theta )$ and ${\cal P}^{(-)}(t,\theta )$ while
Fig.~\ref{fig2} shows a classical sum of the near-side and far-side
intensities.

Note that the linear size of the chaotic billiard ~\cite{5} is only
about twice as bigger as the spatial width of the wave packet at
$t=0$. Therefore, at any moment of time, the width of the wave packet
can not exceed its width at $t=0$ more than by about a factor 2. From
the obtained criterion the interference fringes wash out when the
width of the spreading wave packets begins to exceed its initial value
at $t=0$ by a factor $\geq 2$. It could be of interest to check if the
interference fringes, at fixed moments of time, for chaotic billiard 
would disappear upon increasing the
size of the billiard by a factor of 2-3 keeping the width of the wave
packet at $t=0$ the same as in~\cite{5}.

Our approach does not intend to substitute rigorous numerical
calculations~\cite{9}. Yet, it does show that the complicated
many-body collision problem can be accurately represented by a much
simpler physical picture of a weakly damped ($\beta\ll\hbar\omega$)
quantum rotator. This mapping has enabled us to address a problem of
the quantum-classical transition in complex collisions and evaluate a
characteristic time for such a transition. Remarkably, this simple
picture of a weakly damped coherent rotation has also emerged from the
numerical calculations for the F+HD~\cite{10} and He+H$_2^+$~\cite{11}
chemical reactions. Originally this same physical picture of stable
rotational wave packets was revealed for heavy-ion
collisions~\cite{7,18}. Interestingly, this is in spite of the fact
that a characteristic rotation time of the IC formed in bimolecular
reactions is $\sim$8 orders of magnitude longer than that for the IC
formed in heavy-ion collisions.

The obtained criterion yields $t_{q-cl}\simeq t_{spr}/d$, where
$t_{spr}=\hbar/\beta $ is a characteristic time for the spread of a
wave packet~\cite{6}. For classical macroscopic systems the wave
packet spreading may be associated with the chaotic dynamics and
$t_{spr}$ could be macroscopically large. However, since $d$ is in
$\hbar$ units and, in the macroscopic limit, $\hbar$ vanishes, then
$t_{q-cl}\propto\hbar\to 0$. This provides an alternative to the
explanation in~\cite{2} of the practically instantaneous collapse of
coherent superpositions into classical sum of intensities for
macroscopic systems. Unlike decoherence~\cite{2}, in our approach
``the quantum origins of the classical'' are obtained in the absence
of external noise or coupling to an environment and only due to the
unitary evolution of a pure quantum state.

LB and SK acknowledge financial support from the projects IN--101603
(DGAPA--UNAM) and 43375 (CONACyT).

\end{document}